# **From nanoscale to macroscale: applications of nanotechnology to production of bulk ultra-strong materials.**

 For the twenty years since their discovery, the "Holy Grail" of nanotube research has been to produce them in arbitrarily long lengths. They initially were only produced at micron-scale lengths. After intense research, so far they have still only been made in millimeter to centimeter lengths.

 Still, in the micron-scale samples tested, their tensile strength has been measured to be a maximum of 150 gigapascals (GPa) at a density of only 1.6 g/cm$^3$, 200 times stronger than steel at only 1/5th the weight, for an improvement of 1,000 times in strength-to-weight ratio.

 The big question is can we make or combine the nanotubes to macroscale sizes *while maintaining the strength of the individual nanotubes*? Individual carbon nanotubes and the 2-dimensional form monolayer *graphene* have been measured at micron-scale lengths to have tensile strengths in the range of 100 to 150 GPa, [1], [2], [3]. Carbon nanotubes have been combined, intermingled into bundles and threads for awhile now. These always have significantly lower strength than the individual nanotubes, [4]. However, this is because there were many single nanotubes connected together by weaker van der Waals forces rather than the stronger carbon-carbon molecular bonds that prevail in individual nanotubes. In these cases, with separate nanotubes weakly connected end-to-end, they can just peal apart under tensile load. This is explained here, [5].

  However, some tests of *aligned, arrays* of nanotubes at millimeter length scales also showed significantly lower strength than individual micro-scale nanotubes, [6], [7], [8]. This may be because of the predicted effect of longer nanotubes having more defects and therefore becoming weaker. *If this is the case, then rather than pursuing arbitrarily long nanotubes it may be better to pursue methods of bonding the micro-scale nanotubes at their ends so that their ultra high strength is maintained.  Some possibilities will be suggested in the following pages.*

 **Joining nanotubes to arbitrary lengths.**

 Tying ropes together has been known to create longer ropes whose strength can be 80% to 90% as strong as the component ropes, [9], [10]. Quite key then is that the capability exists to manipulate individual nanotubes at the nanoscale:

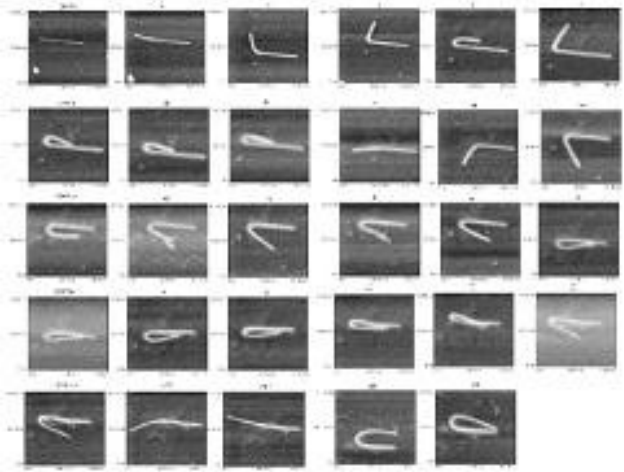

*The stress test: One experiment repeatedly bent a nanotube through contortions to see if it would break. All these modifications were performed by the NanoManipulator, with the user guiding the AFM tip by moving the Phantom force-feedback pen.* [11], [12], [13]

See also, [14], [15], [16]. *Then the suggestion is to tie the nanotubes together using some of the knots known to maintain near the strength of the original ropes. (patent pending.)*

To prevent slipping of the nanotubes under high tensile load we might use them with "nanobuds" along their lengths, [17].

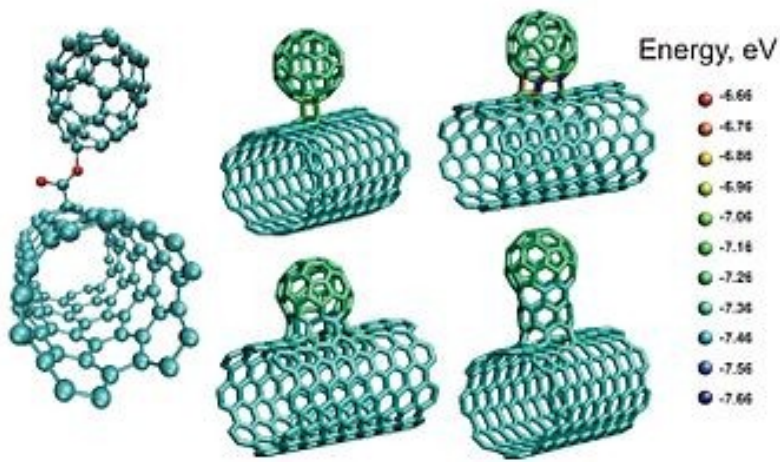

Also, since the nanotubes are quite thin they might be expected to cut into each other when knotted thus weakening the strength of the knot. One possibility might be to fill the portion of nanotube that is to be knotted with water or other fluid to make the nanotube more spongy there, [18].

This method of tying nanotubes together to produce greater lengths has already been proven to work to preserve at least one characteristic of nanotubes, high conductivity:

Energy
Nanotube Cables Hit a Milestone: As Good as Copper.
*Researchers achieve a goal they've been after since the 1980s—the advance could make cars and airplanes lighter, and renewable energy more practical.*
Monday, September 19, 2011 By Katherine Bourzac
http://www.technologyreview.com/energy/38615/

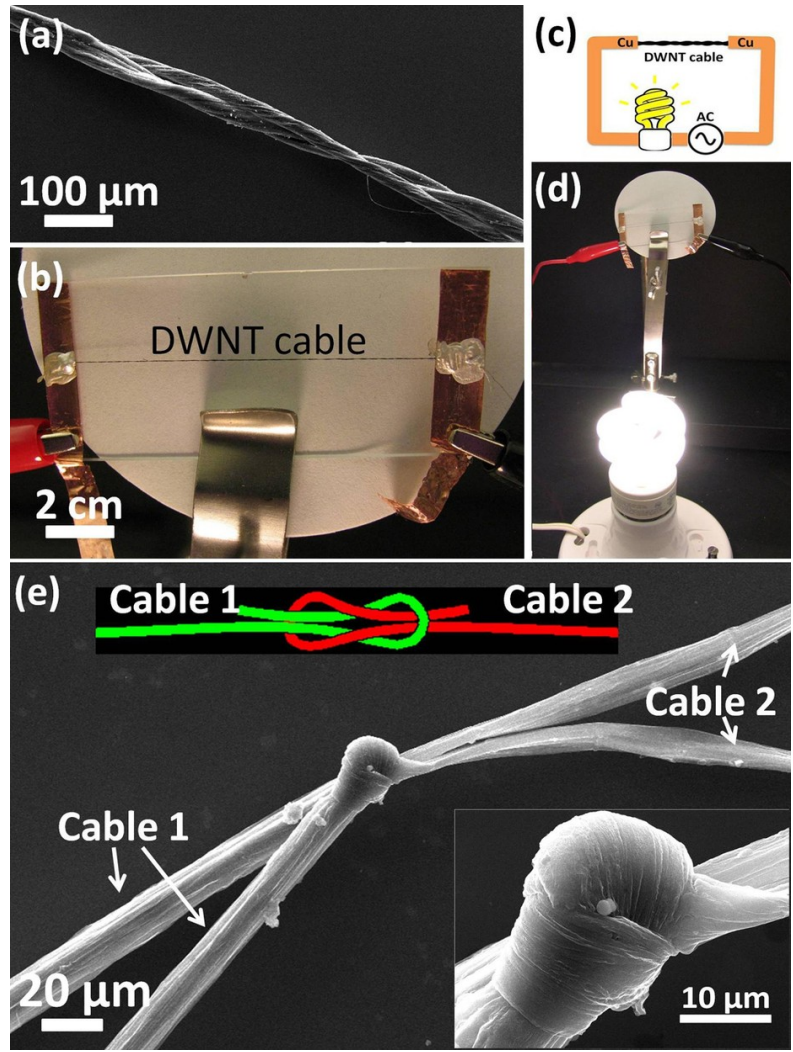

The article describes research by scientists at Rice University who created lightweight electrical cables by mechanically tying together nanotubes.

An alternative method for linking the nanotubes together would be to connect them with nanotube rings:

Ring Closure of Carbon Nanotubes.
Science, Vol. 293, No. 5533, p. 1299-1301, 17 August 2001
*Lightly etched single-walled carbon nanotubes are chemically reacted to form rings. The rings appear to be fully closed as opposed to open coils, as ring-opening reactions did not change the structure of the observed rings. The average diameter of the rings was 540 nanometers with a narrow size distribution.* [19]

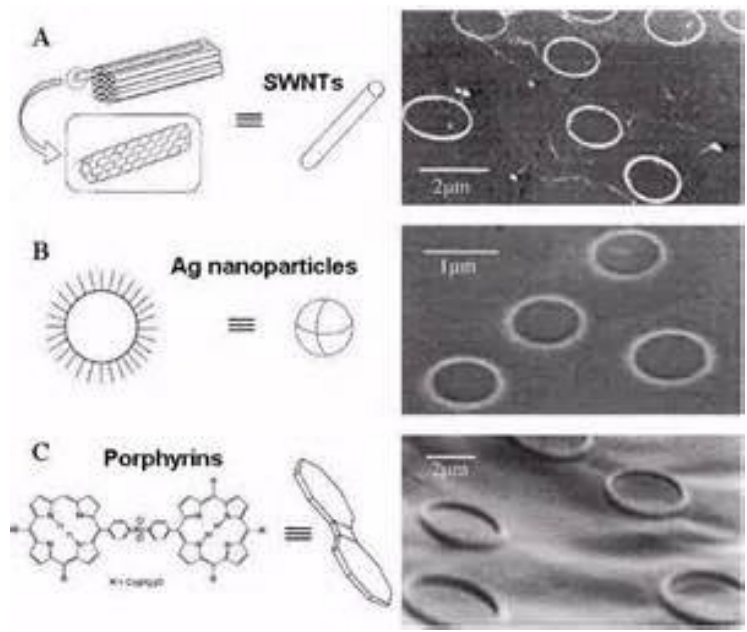

   These are closed rings formed from one or more nanotubes. They are about 540 nm across so several of the aligned nanotubes would have to be fitted into the rings. One question would be how to tighten the rings around the nanotubes once they were fitted into the rings. One possibility might be to apply heating to the rings so that they lengthen then insert the nanotubes inside. Then as the rings cooled they would shrink back to their normal size forming a tight stricture around the nanotubes. As before with the knotting we may have to fill the rings with a fluid so that they are spongy and don't cut into the nanotubes.

   Another method for fitting the carbon nanotubes into the rings would be by using ring-shaped nanotubes of materials that are piezoelectric. Carbon nanotubes are not piezoelectric but nanotubes of many different types of materials have been made, such as boron nitride nanotubes and zinc oxide nanotubes. Nanotubes of both these types are piezoelectric, and they can also be made in the form of nanorings, [20], [21]. Then we could apply electric current to these nanorings to get them to expand, insert the carbon nanotubes, then remove the current to get the nanorings to shrink back to their regular size.

   Note that using the rings as a means of binding the ropes together means you are using frictional effects to get the nanotubes to hold together. Then is this any better than the van der Waals forces holding just intermingled nanotubes together? I believe it can be as long as you make the rings stricture tight enough. But if it is made too tight, this would cut into the nanotube ropes reducing their strength. Then the optimal degree of tightening would have to be found to maintain the greatest strength.
 Interestingly, the method of knotting the nanotubes together or binding them by rings might also be applied to the intermingled bundles, that is, to the case where the nanotubes are of different lengths held together by van der Waals forces. You would note the shortest length of the nanotubes composing a bundle and tie knots around the bundle or bind it with rings at short enough intervals to insure that every nanotube is held tightly with a tie or knot at least once all along the length of the bundle.

Another question that would need to be answered is how binding together a group of equally long nanotubes effects the strength of the nanotubes when the binding is only going around the outer nanotubes. That is, suppose you created a string made from *single* nanotubes bound end-to-end and measured the string's strength.

Then you composed a string by using aligned nanotube arrays that all contained the same number of nanotubes, say 100, and bound these ropes end-to-end with the rings. Would the string composed of the aligned ropes be able to hold 100 times as much as the string composed of individual nanotubes? This is asking a somewhat different question than how knotting weakens the nanotubes. It's asking how strong a composed string will be when a binding can only go around the outer nanotubes composing the string.

Yet another mechanical method for joining the ends together might be to use some nanotubes bent into shapes as clamps. Since nanotubes have such high stiffness they should as clamps be able to hold the ends of aligned arrays of nanotubes together. Again so the clamps don't cut into the nanotubes you might want to have the clamping nanotubes and/or the nanotubes that are being tied to be fluid filled.

A different method of joining individual nanotubes or aligned nanotube arrays end-to-end is suggested by the recent research that created diamond-nanotube composites, [22], [23]. To form the strongest bonds for our purposes, I suggest that the method of creating the strongest nanotubes be used first to create the nanotubes, the arc-discharge method by which the 150 GPa tensile strength nanotubes were made, as in [1]. Then the ends of separate nanotubes or nanotube arrays should be placed on the same diamond seed particle and the high strength microwave CVD method of [4] be used to grow diamond around the ends of both, encasing the tips of each of them inside the diamond thus grown. In order to keep the weight low, you only use a small seed particle and you only grow the diamond large enough to maintain the strong bonds that prevail in individual nanotubes.

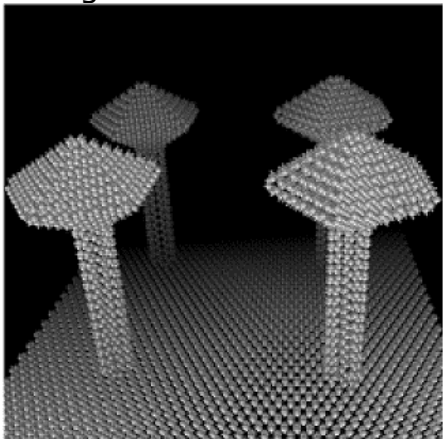

**Figure 3.** Field emitting arrays consisting of diamond pentarparticle/(5,5) nanotube hybrid structures.

Another highly promising method for joining the nanotube ends arises from the surprising effects found by irradiating nanotubes by electron beam:

Reinforcement of single-walled carbon nanotube bundles by intertube bridging.
Nature Materials, 3, p. 153 – 157, March 2004
*During their production, single-walled carbon nanotubes form bundles. Owing to the weak van der Waals interaction that holds them together in the bundle, the tubes can easily slide on each other, resulting in a shear modulus comparable to that of graphite. This low shear modulus is also a major obstacle in the fabrication of macroscopic fibres composed of carbon nanotubes. Here, we have introduced stable links between neighbouring carbon nanotubes within bundles, using moderate electron-beam irradiation inside a transmission electron microscope.* [24]

NEWS & VIEWS
Strong bundles.
Nature Materials, 3, 135-136, March 2004.

*The mechanical properties of nanotube bundles are limited by the sliding of individual nanotubes across each other.*

*Introducing crosslinks between the nanotubes by electron irradiation prevents sliding, and leads to dramatic improvements in strength.* [25]

The researchers noted as had others that intermingled bundles of nanotubes were relatively weak compared to the strength of individual tubes, in this case their measurements being of bending strength. However, after electron beam irradiation the bundles achieved almost 70% of the bending modulus strength of individual nanotubes. A similar effect was seen in [26], [27], [28]. The irradiation produced interconnections between the nanotubes that prevented slipping. Then quite likely this can also be used to combine nanotubes at their ends.
Electron beam irradiation has also been used to attach nanotubes to sensors in scanning electron microscopes for strength testing. One method used was to direct a small amount of hydrocarbons by focused e-beam to weld the nanotubes to the SEM sensor tip. Then this may also work to weld nanotube ends together, [29]. Note that e-beam irradiation can also be used in concert with the tying or ring binding methods to insure no slipping of the nanotubes.

Additionally laser irradiation has been used to connect double-walled nanotubes strands together, [30]. This resulted in longer nanotube strands as strong as the original ones. However, the starting strength of these was low at 335.6 MPa. It needs to be tested if this method can maintain the strength of the original nanotubes at the highest measured strengths of 150 GPa.

Note that these e-beam or laser irradiation methods may also work to produce graphene sheets of large size as well. Currently the 2-dimensional graphene has only been produced in micron-scale sizes, though its strength has been shown to be comparable to that of the highest measured strengths of the nanotubes at 130 GPa, [3]. However, irradiating overlapping graphene sheets on their edges may also allow these to be bonded together.

**Friction-stir welding of nanotube arrays.**
Another method for joining the aligned arrays of nanotubes might be the method friction-stir welding. This method is used to weld metals while maintaining relatively low temperatures. This reduces the damage to the metals and helps to maintain strength. Since this uses relatively low temperatures it may also work to combine the ends of the aligned arrays of nanotubes.

**The Space Elevator.**
Such high strengths in the 100 to 150 GPa range if they can be maintained in the bonded nanotubes are within the range to make the "space elevator" possible.

However, even at such high strengths it is expected the space elevator ribbon would require tapering. Then you would need a means of connecting nanotubes ropes to each other of ever increasing diameter. One possibility for accomplishing this might be by using the "y-shaped" nanotubes, [31]. These are nanotubes that branch off into a Y-shape. If each branch is as strong as the base column then we could attach a base column of one to a branch of another, thereby creating larger and larger diameters.

Using "y-shaped" nanotubes might also be a way to maintain the high strength across connections in general, assuming each branch is as strong as the base, if multiple branches of one are attached to multiple branches of another. To continue this indefinitely, you would need the y-branches to be on both ends of each nanotube.

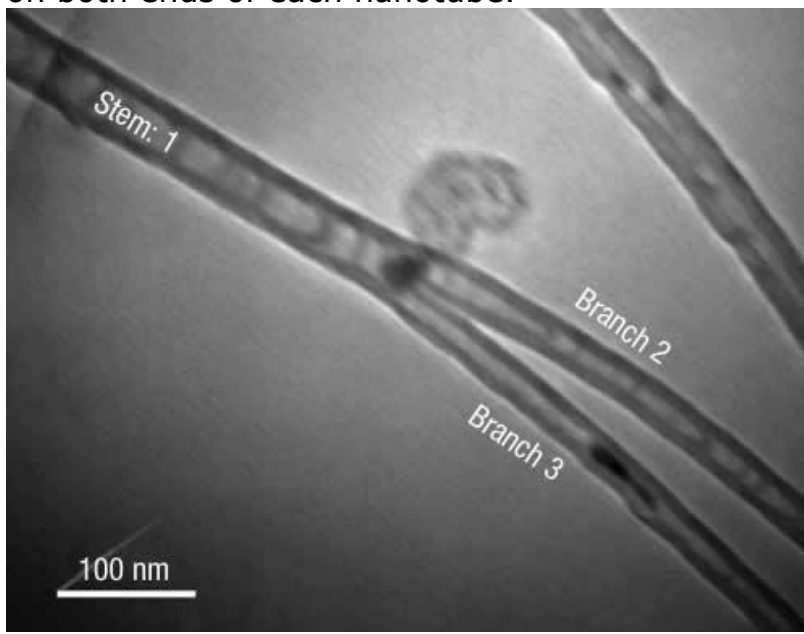

If in general, the connections weakened the strength by some factor we would just use enough branches so that the total strength would be the same as the individual nanotubes. Then if the branches are quite short compared to the base column, the total mass would be just a small fraction larger than that of just the base columns alone, so the strength to weight ratio would be about the same.

In regards to the space elevator, NASA and the SpaceWard Foundation had sponsored a competition with a $1 million prize for a team that can produce a cable material at about double the strength to weight ratio of the strongest commonly used materials now:

Tether Strength Competition.
By the numbers:
Tether Length: 2 m (closed loop)
Tether Weight: 2 g
Breaking Force: 1 ton, 1.5 ton (approx)
Prize Purse: $900k, $1.1M
Best performance to date: 0.72 Ton
Number of Teams: None Yet
Competition Date: February-March, 2009.
http://www.spaceward.org/elevator2010-ts [32]

*I believe both a carbon nanotube cable joined by one the methods described above and a cable made of the new synthetic diamond could each win this competition.*

___________________________________

Robert Clark

Department of Mathematics

Widener University

Chester, PA 19013

610-499-1243

rgc0300@mail.widener.edu

___________________________________